\magnification=1100
\centerline{\bf PATHWAY PARAMETER AND THERMONUCLEAR FUNCTIONS}
\vskip.3cm 
\noindent
\centerline{\bf A.M. Mathai} 
\vskip.1cm
\noindent 
\centerline{Centre for Mathematical Sciences Pala Campus}
\vskip.1cm
\noindent
\centerline{Arunapuram P.O., Palai, Kerala 686 574, India, and}
\vskip.1cm 
\noindent 
\centerline{Department of Mathematics and Statistics, McGill University, Montreal, Canada H3A 2K6} 
\vskip.5cm
\noindent
\centerline{\bf H.J. Haubold} 
\vskip.1cm 
\noindent
\centerline{Office for Outer Space Affairs, United Nations} 
\vskip.1cm 
\noindent 
\centerline{Vienna International Centre, P.O. Box 500, A-1400 Vienna, Austria} 
\vskip.3cm 
\noindent 
{\bf Abstract.} In the theory of thermonuclear reaction rates, analytical evaluation of thermonuclear functions for non-resonant reactions, including cases with cut-off and depletion of the tail of the Maxwell-Boltzmann distribution function were considered in a series of papers by Mathai and Haubold (1988). In the present paper we study more general classes of thermonuclear functions by introducing a pathway parameter $\alpha$, so that when $\alpha \rightarrow 1$ the thermonuclear functions in the Maxwell-Boltzmannian case are recovered. We will also give interpretations for the pathway parameter $\alpha$ in the case of cut-off and in terms of moments.
\vskip.3cm
\noindent 
{\bf 1.\hskip.3cm Thermonuclear Functions}
\vskip.3cm 
The standard thermonuclear function in the Maxwell-Boltzmann case in the theory of nuclear reactions, is given by the following (Critchfield, 1972; Haubold and Mathai, 1985; Mathai and Haubold, 1988):
$$ I_1=\int_0^{\infty}x^{\gamma-1}{\rm e}^{-ax-bx^{-\rho}}{\rm d}x, a>0, b>0, \rho>0,x>0.\eqno(1)$$
Several modifications to this standard thermonuclear function in (1) are considered in Mathai and Haubold (1988), Saxena et al. (2004), and Lissia and Quarati (2005). One such case is the cut-off of the distribution function at point $d$, in which the integral becomes
$$
I_2=\int_0^d x^{\gamma-1}{\rm e}^{-ax-bx^{-\rho}}{\rm d}x, a>0,b>0,\rho>0,d<\infty.\eqno(2)
$$
Another case is depletion of the tail of the distribution function where the integral becomes
$$
I_3=\int_0^{\infty}x^{\gamma-1}{\rm e}^{-ax^{\delta}-bx^{-\rho}}{\rm d}x, a>0,b>0,\rho>0,\delta>0.\eqno(3)
$$
Note that if $\delta=1$ is taken as the standard Maxwell-Boltzmannian behavior, then for $\delta>1$ the right tail will deplete faster and if $\delta<1$ then the depletion will be slower in $I_3$. We consider more general classes of (1), (2), and (3) by replacing ${\rm e}^{-ax^{\delta}}$ by a binomial factor $[1-a(1-\alpha)x^{\delta}]^{{1}\over{1-\alpha}}$ so that the integrand $x^{\gamma-1}[1-a(1-\alpha)x^{\delta}]^{{1}\over{1-\alpha}}$ stays in the generalized type-1 beta family when $\alpha<1$, type-2 beta family when $\alpha>1$ and generalized gamma family when $\alpha\rightarrow 1$. Thus we
consider the general class of integrals
$$
I_{1\alpha}^{(\delta)}=\int_0^{\infty}x^{\gamma-1}[1-a(1-\alpha)x^{\delta}]^{{1}\over{1-\alpha}}{\rm
e}^{-bx^{-\rho}}{\rm d}x,\eqno(4)
$$
for $a>0,\delta>0,b>0,\rho>0$ and
$$
I_{2\alpha}^{(\delta)}=\int_0^dx^{\gamma-1}[1-a(1-\alpha)x^{\delta}]^{{1}\over{1-\alpha}}{\rm
e}^{-bx^{-\rho}}{\rm d}x,\eqno(5)
$$
for $a>0,d>0,\delta>0,b>0,\rho>0$.

For $\alpha<1$ we have extensions of generalized type-1 beta family of integrals in (4) and (5), whereas for $\alpha >1$ we have extensions of generalized type-2 beta family of integrals in (4) and (5). When $\alpha\rightarrow 1$ both of these forms will go to $I_3$ in (3). This consideration is motivated by particularly taking into account Tsallis statistics as developed in Tsallis (1988, 2004), Gell-Mann and Tsallis (2004), and discussed in Cohen (2005).
\vskip.2cm 
The pathway parameter $\alpha$ enables one to go to different functional forms in (4) and (5), thus creating a pathway through these different families of functional forms as shown in Section 2. Also in Section 2 a connection to generalized entropy of order $\alpha$ is given. In Section 3 interpretations for $\alpha$ in terms of the cut-off $d$ in (2) as well as in terms of moments are given. Section 4 evaluates the thermonuclear functions in terms of H-functions. Moments are evaluated in Section 5. Conclusions are drawn in Section 6. For all Sections, concerning generalized entropies of order $\alpha$, see Mathai and Rathie (1975), Gell-Mann and Tsallis (2004), and Mathai and Haubold (2007). 
\vskip.3cm 
\noindent 
{\bf 2.\hskip.3cm Optimization of an Entropy} 
\vskip.3cm 
Let us start with the optimization of the generalized entropy of order $\alpha$,
$$M_{\alpha}(f)=\int_x{{[f(x)]^{2-\alpha}{\rm d}x-1}\over{\alpha-1}},\alpha\ne 1\eqno(6)
$$under the conditions
\vskip.2cm
(i): $f(x)\ge 0$ for all $x$,

(ii): $\int_xf(x){\rm d}x<\infty$, 

(iii): $\int_xx^{\gamma(1-\alpha)}f(x)$ ${\rm d}x=$ fixed for all $f$,

(iv): $\int_xx^{\gamma(1-\alpha)+\delta}f(x){\rm d}x=$ fixed for all $f$, where $\gamma$ and $\delta$ are fixed parameters.
\vskip.2cm
In this discussion we will consider $x$ to be a real scalar variable. For fixed $\alpha$, optimization of $M_{\alpha}(f)$ implies optimization of $\int_x[f(x)]^{2-\alpha}{\rm d}x$ subject to conditions (i) to (iv). If calculus of variation is used then the Euler equation is the following:
$$\eqalignno{{{\partial}\over{\partial
f}}[f^{2-\alpha}&-\lambda_1x^{\gamma(1-\alpha)}f+\lambda_2x^{\gamma(1-\alpha)+\delta}f]=0\Rightarrow\cr
(2-\alpha)f^{1-\alpha}&=\lambda_1x^{\gamma(1-\alpha)}[1-{{\lambda_2}\over{\lambda_1}}x^{\delta}],~\alpha\ne
1,2,\Rightarrow\cr
f_1&=c_1x^{\gamma}[1-a(1-\alpha)x^{\delta}]^{{1}\over{1-\alpha}}&(7)\cr}
$$for ${{\lambda_2}\over{\lambda_1}}=a(1-\alpha)$ for some $a>0,
c=({{\lambda_1}\over{2-\alpha}})^{{1}\over{1-\alpha}}$. Here $c_1$
can act as the normalizing constant when $f$ in (7) is a statistical
density function. 
\vskip.2cm 
For $\alpha<1, a>0, \delta>0$ in (7) we need an additional condition to make (7) a density, namely $1-a(1-\alpha)x^{\delta}>0$ $\Rightarrow 0<x<{{1}\over{[a(1-\alpha)]^{{1}\over{\delta}}}}$ when $x$ is positive. When $\alpha>1$ then writing $1-\alpha=-(\alpha-1)$ we have
$$f_2=c_2x^{\gamma}[1+a(\alpha-1)x^{\delta}]^{-{{1}\over{\alpha-1}}}, \alpha>1, a>0, \delta>0.\eqno(8)
$$When $\alpha\rightarrow 1$, both $f_1$ and $f_2$ go to $f_3$, where
$$f_3=c_3x^{\gamma}{\rm e}^{-ax^{\delta}}, a>0, \delta>0.\eqno(9)
$$In (7), (8) and (9) one can interpret $\delta$ as the rate of depletion of the right tail. If $\delta=1$ is taken as the standard behavior, then for $\delta>1$ the tail is depleted at a faster rate and when $\delta<1$ it is depleted at a slower rate. Then $\delta$ measures the rate of depletion of the right tail. Note that the normalizing constants $c_1,c_2,c_3$ will be different for the three cases $-\infty <\alpha<1, \alpha>1$ and $\alpha <1$. 
\vskip.2cm 
The scalar version of the pathway model of Mathai (2005) is
$$f=c|x|^{\gamma}[1-a(1-\alpha)|x|^{\delta}]^{{\eta}\over{1-\alpha}},
\eta>0, a>0\eqno(10)
$$ for $-\infty<x<\infty$. Our cases in (7),(8), (9) are special cases of (10) for $\eta=1, x>0$. Note that the normalizing constant $c$ in (10) is also different for the three cases $\alpha<1, \alpha>1$, and $\alpha\rightarrow 1$. Observe also that in (8) and (9), $\delta$ could be negative also. In (10) for $\alpha>1$ and $\alpha\rightarrow 1$, $\delta$ could be negative also. 
\vskip.2cm
We look at the normalizing constants $c_1,c_2,c_3$ in (7) to (9) for $x>0$. For $\alpha <1$ make the substitution
$y=a(1-\alpha)x^{\delta}=$. Then ${\rm d}x={{y^{{{1}\over{\delta}}-1}{\rm d}y}\over{\delta[a(1-\alpha)]^{{1}\over{\delta}}}}$ and then
$$\eqalignno{1&=\int_xf_1(x){\rm d}x\cr
&={{c_1}\over{\delta[a(1-\alpha)]^{{\gamma+1}\over{\delta}}}}\int_0^1y^{{{\gamma+1}\over{\delta}}-1}
(1-y)^{{1}\over{1-\alpha}}{\rm d}y\cr
&={{c_1}\over{\delta[a(1-\alpha)]^{{\gamma+1}\over{\delta}}}}
{{\Gamma({{\gamma+1}\over{\delta}})\Gamma({{1}\over{1-\alpha}}
+1)}\over{\Gamma({{1}\over{1-\alpha}}+1+{{\gamma+1}\over{\delta}})}}\cr}$$
evaluating the integral with the help of a type-1 beta integral, for $\gamma+1>0, \alpha <1$. Then
$$c_1=\delta[a(1-\alpha)]^{{\gamma+1}\over{\delta}}
{{\Gamma({{1}\over{1-\alpha}}+1+{{\gamma+1}\over{\delta}})}\over{\Gamma({{\gamma+1}\over{\delta}})
\Gamma({{1}\over{1-\alpha}}+1)}}\eqno(11),
$$for $\alpha<1, \gamma+1>0, \delta>0, a>0$. The $h$-th moment of $x$, denoted by $E(x^h)$, where $E$ denotes the expected value, is then given by
$$
E(x^h)={{1}\over{[a(1-\alpha)]^{{h}\over{\delta}}}}{{\Gamma({{\gamma+1+h}\over{\delta}})}\over{\Gamma({{\gamma+1}\over{\delta}})}}
{{\Gamma({{1}\over{1-\alpha}}+1+{{\gamma+1}\over{\delta}})}\over{\Gamma({{1}\over{1-\alpha}}+1+{{\gamma+1+h}\over{\delta}})}},\eqno(12)
$$for $\alpha<1,\gamma+1+h>0, \delta>0,a>0$. Note that all the $h$-th moments for $h>-1$ exist when $\gamma=0$. Proceeding the same way we have the following $h$-th moment for $\alpha>1$. That is, for $\alpha>1$,
$$E(x^h)={{1}\over{[a(\alpha-1)]^{{h}\over{\delta}}}}{{\Gamma({{\gamma+1+h}\over{\delta}})}\over{\Gamma({{\gamma+1}\over{\delta}})}}
{{\Gamma({{1}\over{\alpha-1}}-{{\gamma+1+h}\over{\delta}})}\over{\Gamma({{1}\over{\alpha-1}}-{{\gamma+1}\over{\delta}})}}\eqno(13)
$$for ${{1}\over{\alpha-1}}-{{\gamma+1}\over{\delta}}>0,
{{1}\over{\alpha-1}}-{{\gamma+1+h}\over{\delta}}>0, \gamma+1+h>0,
\gamma+1>0$ or $-\gamma-1<h<{{\delta}\over{\alpha-1}}-\gamma-1,
\alpha>1$, and the normalizing constant
$$c_2=\delta[a(\alpha-1)]^{{\gamma+1}\over{\delta}}
{{\Gamma({{1}\over{\alpha-1}})}\over{\Gamma({{\gamma+1}\over{\delta}})
\Gamma({{1}\over{\alpha-1}}-{{\gamma+1}\over{\delta}})}}\eqno(14)
$$for ${{1}\over{\alpha-1}}-{{\gamma+1}\over{\delta}}>0, \gamma+1>0,
\delta>0, \alpha>1$. 
When $\alpha\rightarrow 1$ then $c_1$ and $c_2$ reduce to
$$c_3={{\delta[a^{{\gamma}\over{\delta}}]}\over{\Gamma({{\gamma+1}\over{\delta}})}},
\gamma+1>0\eqno(15)
$$and for $\alpha\rightarrow 1$,
$$E(x^h)={{1}\over{a^{{h}\over{\delta}}}}
{{\Gamma({{\gamma+1+h}\over{\delta}})}\over{\Gamma({{\gamma+1}\over{\delta}})}},\eqno(16)
$$for $\alpha\rightarrow 1, \gamma+1+h>0, \gamma+1>0, a>0,\delta>0$.
Proceeding the same way one can compute $c$ and $E(|x|^h)$ in (10) for the three cases $\alpha<1,\alpha>1$ and $\alpha\rightarrow 1$. Since the procedure is parallel we will not give the details here.
\vskip.3cm 
\noindent 
{\bf 3.\hskip.3cm Pathway Parameter $\alpha$} 
\vskip.3cm
In the above formulation the parameter $\alpha$ creates a pathway to three different functional forms. For $\alpha<1$ the pathway density of (10) stays in the generalized type-1 beta family of distributions. As $\alpha$ moves closer and closer to $1$ we move continuously to a generalized gamma or exponential type family of densities. As $\alpha$ goes above $1$ we move from the generalized gamma type family to a generalized type-2 beta form. Thus the pathway parameter 
$\alpha$ takes one to three different functional forms. This is the {\it distributional pathway}. 
\vskip.2cm 
Let us see what happens to the generalized entropy of order $\alpha$ in (6), when $\alpha$ moves from $-\infty$ to 
$\infty$. When $\alpha$ approaches $1$, $M_{\alpha}(f)$ comes closer to Shannon's entropy and finally
$$\lim_{\alpha\rightarrow 1}M_{\alpha}(f)=-\int f(x)\ln f(x){\rm
d}x=S(f)= {\hbox{ Shannon entropy}}.
$$Then when $\alpha$ increases from $1$, $M_{\alpha}(f)$ moves away from Shannon entropy to the form in (6). Thus $\alpha$ creates a pathway for the generalized entropy in (6) also. This is the {\it entropic pathway}. 
\vskip.2cm 
The {\it differential pathway} provided by the pathway parameter $\alpha$ can be seen from the following: As an example, let us start from (2). Let
$$\eqalignno{g(x)&={{f_1(x)}\over{c_1}}=x^{\gamma}[1-a(1-\alpha)x^{\delta}]^{{1}\over{1-\alpha}},
x>0\cr {{{\rm d}}\over{{\rm
d}x}}g(x)&={{\gamma}\over{x}}g(x)-a\delta
x^{\delta-1+(1-\alpha)\gamma}[g(x)]^{\alpha}\cr
&=-a[g(x)]^{\alpha}{\hbox{ ÿfor ÿ}}\gamma=0,\delta=1.&(17)\cr}
$$This is the power law. The differential equation satisfied by $g_3={{f_3}\over{c_3}}$ of (9) is given by
$$\eqalignno{{{{\rm d}}\over{{\rm
d}x}}g_3(x)&=({{\gamma}\over{x}}-a\delta x^{\delta-1})g_3(x)\cr
&=-ag_3(x) {\hbox{ ÿfor ÿ}}\gamma=0, \delta=1.&(18)\cr}
$$Thus when $\alpha$ moves to $1$ the differential pathway is from the power law in (17) to the Maxwell-Boltzmann law in (18). 
\vskip.3cm
\noindent 
{\bf 3.1.\hskip.3cm An interpretation of the pathway parameter $\alpha$} 
\vskip.3cm 
Let us start with the pathway density in (7). For this to remain a density we need the condition $1-a(1-\alpha)x^{\delta}>0$ or when $x$ is positive then $0<x<{{1}\over{[a(1-\alpha)]^{{1}\over{\delta}}}}, \alpha <1$, or if we have the model in (5) then
$$-{{1}\over{[a(1-\alpha)]^{{1}\over{\delta}}}}<x<{{1}\over{[a(1-\alpha)]^{{1}\over{\delta}}}},
\alpha <1.$$
Outside this range the density is zero. Thus for $x>0$ the right tail of the distribution function is cut-off at
${{1}\over{[a(1-\alpha)]^{{1}\over{\delta}}}}$. If $d$ is this cut-off on the right then
$$d={{1}\over{[a(1-\alpha)]^{{1}\over{\delta}}}}\Rightarrow \alpha
=1-{{1}\over{a d^{\delta}}}\hbox{ for }\alpha <1.\eqno(19)
$$As $\alpha$ moves closer to $1$ then the cut-off $d$ moves farther out and eventually when $\alpha\rightarrow 1$ then $d\rightarrow\infty$. In this case $\alpha$ is computed easily from the cut-off. 
\vskip.3cm
\noindent 
{\bf 4.\hskip.3cm Evaluation of the Generalized Thermonuclear Functions} 
\vskip.3cm 
Evaluation of $I_{1\alpha}^{(\delta)}$ and $I_{2\alpha}^{(\delta)}$ in (4) and (5) can be done by using the convolution property of Mellin transforms by writing in the form
$$\int_0^{\infty}{{1}\over{x}}f_1(x)f_2({{u}\over{x}}){\rm d}x
$$where
$$f_1(x)=x^{\gamma}[1-a(1-\alpha)x^{\delta}]^{{1}\over{1-\alpha}}{\hbox{ and
}}f_2(x)={\rm e}^{-x^{\rho}}\eqno(20)
$$so that $f_2({{u}\over{x}})={\rm e}^{-u^{\rho}x^{-\rho}}$ and that
$b^{{1}\over{\rho}}=u$. For $\alpha>1$ we write
$f_1(x)=x^{\gamma}[1+a(\alpha-1)x^{\delta}]^{-{{1}\over{\alpha-1}}}$
and for $\alpha<1$ we write
$f_1(x)=x^{\gamma}[1-a(1-\alpha)x^{\delta}]^{{1}\over{1-\alpha}}$.
According to the convolution property of Mellin transforms, the Mellin transform of $I_{1\alpha}^{(\delta)}$ is the product of the Mellin transforms of $f_1$ and $f_2$, denoted by $M_{f_1}(s)$ and $M_{f_2}(s)$ respectively, where for $\alpha >1$, taking the more generalized forms of $f_1$, we have for $\delta >0$
$$\eqalignno{M_{f_1}(s)&=\int_0^{\infty}x^{\gamma+s-1}[1+a(\alpha-1)x^{\delta}]^{-{{1}\over{\alpha-1}}}{\rm
d}x\cr
&={{\Gamma({{(\gamma+s)}\over{\delta}})\Gamma({{1}\over{\alpha-1}}-{{(\gamma+s)}\over{\delta}})}
\over{[a(\alpha -1)]^{{(\gamma
+s)}\over{\delta}}\Gamma({{1}\over{\alpha-1}})}}, \alpha>1,\gamma
+s>0,{{1}\over{\alpha-1}}-{{(\gamma +s)}\over{\delta}}>0.\cr
M_{f_2}(s)&=\int_0^{\infty}x^{s-1}{\rm e}^{-x^{\rho}}{\rm
d}x={{1}\over{\rho}}\Gamma({{s}\over{\rho}}), s>0.\cr}$$
Thus for
$\alpha >1$,
$$\eqalignno{M_{f_1}(s)M_{f_2}(s)&={{1}\over{\delta\rho
[a(\alpha-1)]^{{(\gamma +s)}\over{\delta}}}}\cr
&\times
{{\Gamma({{s}\over{\rho}})\Gamma({{(\gamma +s)}\over{\delta}})
\Gamma({{1}\over{\alpha-1}}-{{(\gamma
+s)}\over{\delta}})}\over{\Gamma({{1}\over{\alpha-1}})}},\cr}$$for
$s>0, \gamma +s>0$,$\rho>0, \delta>0, a>0$,
${{1}\over{\alpha-1}}-{{(\gamma +s)}\over{\delta}} >0, \alpha >1$.
Then the integral in $I_{1\alpha}^{(\delta)}$ is available by taking the inverse Mellin transform. That is,
$$\eqalignno{I_{1\alpha}^{(\delta)}&={{1}\over{\delta\rho
[a(\alpha-1)]^{{\gamma}\over{\delta}}}}
{{1}\over{\Gamma({{1}\over{\alpha-1}})}}\cr &\times {{1}\over{2 \pi
i}}\int_{c-i\infty}^{c+i\infty}\Gamma({{s}\over{\rho}})\Gamma({{(\gamma+s)}\over{\delta}})
\Gamma({{1}\over{\alpha-1}}-{{(\gamma+s)}\over{\delta}})(b^{{1}\over{\rho}}[a(\alpha-1)]^{{1}\over{\delta}})^{-s}{\rm
d}s\cr &={{1}\over{\delta\rho
[a(\alpha-1)]^{{\gamma}\over{\delta}}}}{{1}\over{\Gamma({{1}\over{\alpha-1}})}}
H_{1,2}^{2,1}\left[b^{{1}\over{\rho}}[a(\alpha-1)]^{{1}\over{\delta}}\big\vert_{(0,{{1}\over{\rho}}),
({{\gamma}\over{\delta}},
{{1}\over{\delta}})}^{({{(\alpha-2)}\over{(\alpha-1)}}+{{\gamma}\over{\delta}},{{1}\over{\delta}})}\right].\cr
&={{1}\over{\delta\rho
a^{{\gamma}\over{\delta}}}}H_{0,2}^{2,0}\left[a^{{1}\over{\delta}}b^{{1}\over{\rho}}\big\vert_{(0,{{1}\over{\rho}}),
({{\gamma}\over{\delta}}, {{1}\over{\delta}})}\right], {\hbox{ for
}}\alpha\rightarrow 1&(21)\cr}
$$For the definition, theory and applications of H-functions see Mathai and Saxena (1978). Note that in $M_{f_1}(s)$ we could have considered the form with $x^{-\delta}, \delta>0$ instead of $x^{\delta}, \delta>0$. Mathematically, both integrals are tractable. For $x^{\delta}$ with $\delta=1$, we have the extension of the Maxwell-Boltzmann case ${\rm e}^{-ax}$. 
\vskip.2cm 
For $\alpha <1$ we may evaluate the pathway integral corresponding to
$I_{2\alpha}^{(\delta)}$, namely,
$$I_{2\alpha}^{(\delta)}=\int_0^dx^{\gamma-1}[1-a(1-\alpha)x^{\delta}]^{{1}\over{1-\alpha}}{\rm
e}^{-bx^{-\rho}}{\rm d}x, \alpha <1
$$where for $\alpha <1$,
$$f_1(x)=x^{\gamma}[1-a(1-\alpha)x^{\delta}]^{{1}\over{1-\alpha}}{\rm
d}x
$$for $d={{1}\over{[a(1-\alpha)]^{{1}\over{\delta}}}}$ or
$0<x<{{1}\over{[a(1-\alpha)]^{{1}\over{\delta}}}}$ and $f_1(x)=0$
elsewhere, and $f_2(x)={\rm e}^{-x^{\rho}}, 0<x<\infty$. 
Then
$$\eqalignno{M_{f_1}(s)&=\int_0^dx^{\gamma+s-1}[1-a(1-\alpha)x^{\delta}]^{{1}\over{1-\alpha}},
~d={{1}\over{[a(1-\alpha)]^{{1}\over{\delta}}}}, \alpha<1\cr
&={{1}\over{\delta[a(1-\alpha)]^{{(\gamma +s)}\over{\delta}}}}
{{\Gamma({{(\gamma+s)}\over{\delta}})\Gamma({{1}\over{1-\alpha}}+1)}\over{\Gamma({{1}\over{1-\alpha}}
+1+{{(\gamma+s)}\over{\delta}})}}, \gamma+s>0, \alpha <1\cr
M_{f_2}(s)&=\int_0^{\infty}x^{s-1}{\rm e}^{-x^{\rho}}{\rm
d}x={{1}\over{\rho}}\Gamma({{s}\over{\rho}}), s>0, \rho>0.\cr}
$$Hence,
$$\eqalignno{I_{2\alpha}^{(\delta)}&={{1}\over{\delta\rho}}\Gamma({{s}\over{\rho}}){{1}\over{[a(1-\alpha)]^{{\gamma}\over{\delta}}}}
{{1}\over{2\pi
i}}\int_{c-i\infty}^{c+i\infty}{{\Gamma({{(\gamma+s)}\over{\delta}})\Gamma({{1}\over{1-\alpha}}+1)}
\over{\Gamma({{1}\over{1-\alpha}}+1+{{(\gamma+s)}\over{\delta}})}}(b^{{1}\over{\rho}}[a(1-\alpha)]^{{1}\over{\delta}})^{-s}{\rm
d}s\cr
&={{\Gamma({{1}\over{1-\alpha}}+1)}\over{\rho\delta[a(1-\alpha)]^{{\gamma}\over{\delta}}}}
H_{1,2}^{2,1}\left[b^{{1}\over{\rho}}[a(1-\alpha)]^{{1}\over{\delta}}\big\vert_{(0,{{1}\over{\rho}}),
({{\gamma}\over{\delta}},{{1}\over{\delta}})}^{({{1}\over{1-\alpha}}+1+{{\gamma}\over{\delta}},{{1}\over{\delta}})}\right]\cr
&={{1}\over{\delta\rho
a^{{\gamma}\over{\delta}}}}H_{0,2}^{2,0}\left[b^{{1}\over{\rho}}a^{{1}\over{\delta}}\big\vert_{(0,{{1}\over{\rho}}),
({{\gamma}\over{\delta}}, {{1}\over{\delta}})}\right], {\hbox{ for
}}\alpha\rightarrow 1.\cr}
$$For $\delta=1$ we have the extension of the Maxwell-Boltzmann case of ${\rm e}^{-ax}$ in $f_1(x)$. 
\vskip.3cm 
\noindent 
{\bf 5.\hskip.3cm An Evaluation of $\alpha$} 
\vskip.3cm 
In this connection let us evaluate the $h$-th moment in the density
$$f_1(x)=c_1x^{\gamma}[1+a(\alpha-1)x^{\delta}]^{-{{1}\over{\alpha-1}}},
\alpha >1, \delta >0, a>0, x>0
$$where $c_1$ is the normalizing constant. Integration gives the following, denoting the $h$-th moment by $E(x^h)$.
$$E(x^h)={{1}\over{[a(\alpha-1)]^{{h}\over{\delta}}}}{{\Gamma({{(\gamma+h+1)}\over{\delta}})}
\over{\Gamma({{(\gamma+1)}\over{\delta}})}}
{{\Gamma({{1}\over{\alpha-1}}-{{(\gamma+h+1)}\over{\delta}})}\over{\Gamma({{1}\over{\alpha-1}}-{{(\gamma+1)}\over{\delta}})}},
$$for $\alpha >1, \delta>0, \gamma +h+1>0$,
${{1}\over{\alpha-1}}-{{(\gamma+h+1)}\over{\delta}}>0, \gamma +1>0$.
When ${{h}\over{\delta}}=m, m=1,2,...$ or when $h=\delta m, m=1,...$
we have for $m=1$,
$$E(x^{\delta})={{1}\over{a(\alpha-1)}}
{{(\gamma+1)}\over{\delta}}{{1}\over{{{1}\over{\alpha-1}}-{{(\gamma+1)}\over{\delta}}-1}}
$$which gives
$$\alpha={{(\gamma+1+2\delta)}\over{(\gamma+1+\delta)}}-{{(\gamma+1)}\over{(\gamma+1+\delta)aE(x^{\delta})}}
$$for $1<\alpha<1+{{\delta}\over{\gamma+1+\delta}}$. Similarly one
can evaluate $\alpha$ for $\alpha <1$ also from $E(x^{\delta})$. In this case
$$\alpha =1+{{1}\over{(\delta+\gamma+1)}}[\delta-{{(\gamma+1)}\over{aE(x^{\delta})}}]$$
for $\gamma+1>a\delta E(x^{\delta})$.
\vskip.3cm
\noindent
{\bf 6.\hskip.3cm Conclusions}
\vskip.3cm
In the field of stellar, cosmological, and controlled fusion, for example, the core of the Sun is considered as the gravitationally stabilized solar fusion reactor. The probability for a thermonuclear reaction to occur in the solar fusion plasma depends mainly on two factors. One of them is the velocity distribution of the particles in the plasma and is usually given by the Maxwell-Boltzmann distribution of Boltzmann-Gibbs statistical mechanics. The other factor is the particle reaction cross-section that contains the dominating quantum mechanical tunneling probability through a Coulomb barrier, called Gamow factor. Particle reactions in the hot solar fusion plasma will occur near energies where the product of velocity distribution and tunneling probability is a maximum. The product of velocity distribution function and penetration factor is producing the Gamow peak. Mathematically, the Gamow peak is a thermonuclear function. In case of taking into consideration electron screening of reactions in the hot fusion plasma, the Coulomb potential may change to a Yukawa-like potential. Taking into account correlations and long-range forces in the plasma, the Maxwell-Boltzmann distribution may show deviations covered by the distribution predicted by Tsallis statistics in terms of cut-off or depletion of the high-velocity tail of the distribution function. In this paper, closed-form representations have been derived for thermonuclear functions, thus for the Gamow peak, for Boltzmann-Gibbs and Tsallis statistics. For this purpose, a generalized entropy of order $\alpha$ and the distribution function, emanating from optimizing this entropy, has been considered. The case $\alpha = 1$ recovers the Maxwell-Boltzmannian case. This general case is characterized by moving cut-off, respectively the upper integration limit of the thermonuclear function to infinity. The closed-form representations of thermonuclear functions are achieved by using generalized hypergeometric functions of Fox or H-functions.       
\vskip.5cm 
\noindent 
\centerline{{\bf Acknowledgment}} 
\vskip.3cm 
The authors would like to thank the Department of Science and Technology, Government of India, New Delhi, for the financial assistance under project No.
SR/S4/MS:287/05. 
\vskip.3cm 
\noindent 
\centerline{{\bf References}}
\vskip.3cm 
\noindent
Cohen, E.G.D. (2005). Boltzmann and Einstein: statistics and dynamics - an unsolved problem, {\it Pramana}, {\bf 64}, 635-643.
\vskip.2cm
\noindent
Critchfield, C.L. (1972). Analytic forms of thermonuclear function, in {\it Cosmology, Fusion and Other Matters: George Gamow Memorial Volume}, Ed. F. Reines, University of Colorado Press, Colorado, pp. 186-191.
\vskip.2cm
\noindent
Gell-Mann, M. and Tsallis, C. (Eds.). (2004). {\it Nonextensive Entropy: Interdisciplinary Applications}, Oxford University Press, New York.
\vskip.2cm
\noindent
Haubold, H.J. and Mathai, A.M. (1985). The Maxwell-Boltzmannian approach to the nuclear reaction rate theory, {\it Reports on Progress in Physics}, {\bf 33}, 623-644.
\vskip.2cm
\noindent
Lissia, M. and Quarati, P. (2005). Nuclear astrophysical plasmas: ion distribution functions and fusion rates, {\it europhysics news}, {\bf 36}, 211-214. 
\vskip.2cm
\noindent 
Mathai, A.M. (2005). A pathway to matrix-variate gamma and normal densities, {\it Linear Algebra and Its Applications}, {\bf 396}, 317-328. 
\vskip.2cm 
\noindent 
Mathai, A.M. and Haubold, H.J. (1988). {\it Modern Problems in Nuclear and Neutrino Astrophysics}, Akademie-Verlag, Berlin. 
\vskip.2cm
\noindent 
Mathai, A.M. and Haubold, H.J. (2007). Pathway model, superstatistics, Tsallis statistics and a generalized measure of entropy, {\it Physica A}, {\bf 375}, 110-122. 
\vskip.2cm 
\noindent 
Mathai, A.M. and Rathie, P.N. (1975). {\it Basic Concepts in Information Theory and Statistics: Axiomatic Foundations and Applications}, Wiley Eastern, New Delhi and Wiley Halsted, New York.
\vskip.2cm 
\noindent 
Mathai, A.M. and Saxena, R.K. (1978). {\it The H-function with Applications in Statistics and Other Disciplines}, Wiley Eastern, New Delhi and Wiley Halsted, New York.
\vskip.2cm
\noindent
Saxena, R.K., Mathai, A.M., and Haubold, H.J. (2004). Astrophysical thermonuclear functions for Boltzmann-Gibbs statistics and Tsallis statistics, {\it Physica A}, {\bf 344}, 649-656.
\vskip.2cm
\noindent
Tsallis, C. (1988). Possible generalization of Boltzmann-Gibbs statistics, {\it Journal of Statistical Physics}, {\bf 52}, 479-487.
\vskip.2cm
\noindent
Tsallis, C. (2004). What should a statistical mechanics satisfy to reflect nature?, {\it Physica D}, {\bf 193}, 3-34.
\end